\documentclass[12pt]{article}
\usepackage{amsmath,amsthm}
\usepackage{graphicx,psfrag,epsf}
\usepackage{enumerate}
\usepackage[plain,noend]{algorithm2e}
\usepackage{algorithmic}
\usepackage{natbib}
\usepackage{subcaption}
\usepackage{url} 
\usepackage[colorlinks]{hyperref}
\hypersetup{
    colorlinks,
    linkcolor={red},
    filecolor={red}, 
    urlcolor={blue},
    citecolor={blue}
}

\newcommand*{\email}[1]{%
    \normalsize\href{mailto:#1}{#1}\par
    }

\newcommand{\blind}{1}

\def\cD{{\cal D}}

\def\beq#1\eeq{\begin{equation}#1\end{equation}}
\def\baa#1\eaa{\begin{eqnarray}#1\end{eqnarray}}
\def\bal#1\eal{\begin{align}#1\end{align}}

\newtheorem{theorem}{Theorem}

\begin{document}

\def\spacingset#1{\renewcommand{\baselinestretch}%
{#1}\small\normalsize} \spacingset{1}


\if1\blind
{
  \title{\bf Confidence Band Estimation for Survival Random Forests}
  \author{Sarah Formentini \thanks{The first two authors contributes equally} \thanks{\email{sarahef2@illinois.edu}}\hspace{.2cm}\\
    Department of Statistics, University of Illinois\\
    and \\
    Wei Liang $^*$\thanks{\email{wliang@xmu.edu.cn}}\\
    School of Mathematical Sciences, Xiamen University\\
    and\\
    Ruoqing Zhu \thanks{\email{rqzhu@illinois.edu}}\\
    Department of Statistics, University of Illinois}
  \maketitle
} \fi

\if0\blind
{
  \bigskip
  \bigskip
  \bigskip
  \begin{center}
    {\LARGE\bf Title}
\end{center}
  \medskip
} \fi

\bigskip
\begin{abstract}
Survival random forest is a popular machine learning tool for modeling censored survival data. However, there is currently no statistically valid and computationally feasible approach for estimating its confidence band. This paper proposes an unbiased confidence band estimation by extending recent developments in infinite-order incomplete U-statistics. The idea is to estimate the variance-covariance matrix of the cumulative hazard function prediction on a grid of time points. We then generate the confidence band by viewing the cumulative hazard function estimation as a Gaussian process whose distribution can be approximated through simulation. This approach is computationally easy to implement when the subsampling size of a tree is no larger than half of the total training sample size. Numerical studies show that our proposed method accurately estimates the confidence band and achieves desired coverage rate. We apply this method to veterans' administration lung cancer data.
\end{abstract}

\noindent%
\vfill

\newpage
\spacingset{1} 

\section{Introduction}\label{sec:intro}

First introduced by \cite{Breiman2001}, a random forest is a non-parametric, tree-based, bagging ensemble model. It generates predictions based on large numbers of trees built with random mechanics such as sub-sampling and random feature selection. While the original model was proposed for regression or classification problems, it has been extend to survival settings \citep{hothorn2006survival, ishwaran2008random, zhu2012recursively} and found great success in biomedical studies. It is widely used in genetic studies and personalized medicine \citep{Shi2019, Takashima2020, Chen2010, Song2019, Zhang2019}. While forests effectively produce predictions for different tasks, there is minimal literature on their statistical inferences. To the best of our knowledge, no existing method provides a valid confidence band for survival function predictions from a random forest.

Previous works established variance estimations and normality results for regression forests \citep{Mentch2016, Wager2014, Wager2018, zhou2021v, Peng2021}. Two popular approaches have been considered in the literature for estimating the variance of a random forest. By utilizing the bootstrapping mechanism, \cite{sexton2009standard} and \cite{athey2019generalized} propose to use Jackknife and infinitesimal Jackknife to estimate the variance. Another popular approach by \cite{Mentch2016} views random forests as a U-statistic, obtained through subbagging, i.e., subsample aggregating \citep{geurts2006extremely} instead of bootstrap aggregating. This approach permits the estimation of its variance using the U-statistics theory. Under suitable conditions, the leading term in the Hoeffding decomposition \citep{Hoeffding1948} of the U-statistic variance estimation dominates, promising the estimation in an incomplete case \citep{Lee1990U-statistics:Practice}, i.e., with a finite number of trees. These methods required that the subsampling size $k$ is of order $o(n^{1/2})$, while \cite{peng2019asymptotic} relax the condition to $o(n)$ under further assumptions. However, all existing methods suffer in practice when a large subsampling size $k$ is used to fit each tree, typically in the order of $O(n)$. The variance estimation is significantly biased if only the leading term in the Hoeffding decomposition is used. Furthermore, some methods are computationally complex, and modifications must be made \citep{zhou2021v}. This entire issue is further complicated in survival analysis because the survival function needs to be estimated on the domain of the entire survival time, which means dealing with a U-process rather than a U-statistic. A small subsampling size is usually not preferred for random survival forests since the tree structure is too shallow. There is no currently valid approach to this topic due to these theoretical and computational challenges.

Recently, \cite{xu2022variance} proposed an unbiased variance estimation of U-statistics in the regression setting of $k \leq n / 2$. They view the entire Hoeffding decomposition as the target of estimation, which removes the bias issue when the leading term no longer dominates. They utilize the conditional variance formula to convert the estimation to a computationally feasible quantity. This promising approach can be adapted to many random forest approaches and potentially to any subbagging estimators. This paper makes the first attempt to develop a confidence band estimation of random survival forests. Our primary goal is to provide an unbiased estimate of the covariance matrix given a grid of time points and then numerically obtain the critical value that achieves the desired coverage rate.

\section{Methodology}\label{sec:method}
\subsection{Random Forest Notation}
Following standard notation in survival analysis, let $\cD_n=\lbrace Z_i=(X_i,\,Y_i,\,\delta_i)\rbrace_{i=1}^n$ be a set of $n$ i.i.d. copies of the $p$-dimensional predictor variables, observed survival times, and censoring indicators sampled from the population.
$Y_i=\min(T_i,C_i)$ is the observed survival time, $C_i$ is the censoring time for individual $i$, and $\delta_i=I(T_i\leq C_i)$ is the censoring indicator. We assume $T$, the actual survival time, follows a conditional distribution $F(t\,\mid \,x)=\text{pr}(T\leq t\mid  x=X)$. $S(t\,\mid \,x)=1-F(t\,\mid \,x)$ is the survival function, $\Lambda(t\,\mid \,x)=-\log\lbrace S(t\,\mid \,x)\rbrace$ is the cumulative hazard function, and  $\lambda(t\,\mid \,x)=\text{d}\,\Lambda(t\,\mid \,x)/\text{d}t$ is the hazard function.  The censoring time $C$ follows the conditional distribution $G(t\,\mid \,x)=\text{pr}(C\leq t\,\mid \,X)$, where a non-informative censoring mechanism, $T$ independent of censoring $C$ given $X$, is assumed.

Random forest models use a random kernel (tree) function instead of a deterministic kernel function. For the sake of estimating the variance of U-statistics, the analysis of random U-statistic can reasonably boil down to the classical framework of non-random U-statistic, while the only requirement is to use a large number of trees \citep{Mentch2016}.
Each tree provides a hazard or cumulative hazard function estimations at any target point. Hence, we can view a tree prediction at a target point $x_0$ as $ h (t\,\mid \,x_0;\,Z_{1}, \ldots, Z_{k} )$, with $t$ ranging in the domain of the survival time. Suppose $S_{j}=(Z_{j_1}, \ldots, Z_{j_k})$ is the $j$th subsample of size $k$ from the training set. Without the risk of ambiguity, we write
\[ h\left(t\,\mid \,x_0;\,Z_{j_1},\,\ldots,\,Z_{j_k}\right)=h(t;\,S_{j}). \]

A survival random forest model can be viewed as a U-statistic where
\bal
U(t) = \binom{n}{k}^{-1} \sum_{{j_1}<{j_2}<\ldots<{j_k}} h\left(t\,\mid \,x_0;\,Z_{j_1},\,\ldots,\,Z_{j_k}\right)= \binom{n}{k}^{-1} \sum_{S_j\subset\cD_n} h(t;\,S_{j}).\label{eq:Un}
\eal
Unlike traditional U-statistics, the sub-sampling size $k$ usually grows with $n$ in a random survival forest, and there are too many sub-samples to exhaust them all. Hence, a random forest model usually fits $B$ trees, where the pre-specified $B$ is reasonably large. This formulation belongs to the class of incomplete U-statistics. In the following, we will first focus on estimating the variance of $U(t)$ based on its complete version, then extend the method to the incomplete version.

\subsection{Random Forests as Complete U-Statistics}\label{sec:regression}

For a fixed time point $t_0$, the Hoeffding decomposition \citep{Hoeffding1948} provides a nice way to estimate the variance of an order-$k$ complete U-statistics as
\bal
\text{var}\big\lbrace U(t_0)\big\rbrace  = \sum_{d = 1}^k \binom{n}{k}^{-1} \binom{k}{d} \binom{n-k}{k-d}\,\, \xi_{d,\,k}^2(t_0), \label{eq:varofu}
\eal
where $\xi_{d,\,k}^2(t_0)$ is the covariance of two trees $h(t_0;\,S_1)$ and $h(t_0;\,S_2)$ with $\mid S_1\cap S_2\mid = d$ and $\mid\cdot\mid$ is the cardinality of a set. In other words, $S_1$ and $S_2$ share $d$ number of common observations  $Z_{(1:d)}=(Z_1,\,\ldots,\,Z_d)$, and
\[\xi_{d,\,k}^2(t_0) = \text{cov} \big\lbrace  h(t_0;\,S_1),\, h(t_0;\,S_2) \big\rbrace
 = \text{var}\big[ \text{E} \big\lbrace h(t_0;\,S) \mid  Z_{(1:d)} \big\rbrace \big].\]

However, existing methods for estimating the variance of $U(t_0)$ rely mainly on $k$ being small and sometimes also require that $\xi_{1,\,k}^2(t_0)$ is non-degenerate. In that case, the first summand on the right side of \eqref{eq:varofu} dominates all remaining terms. \cite{Mentch2016} developed methods for regression random forests relying on this dominance under certain regularity conditions. \cite{zhou2021v} further extended the results by placing regularity conditions on the ratio of $\xi_{d,\,k}^2(t_0)$ over $\xi_{1,\,k}^2(t_0)$. However, these methods usually suffer when $k$ is large since \eqref{eq:varofu} will be mainly determined by terms $\xi_{d,\,k}^2(t_0)$ with large $d$. Only estimating the first term $\xi_{1,\,k}^2(t_0)$ results in a large bias.

We adopt the strategy developed by \citet{xu2022variance} to create a method without this weakness. The essential idea is to decompose $\xi_{d,\,k}^2(t_0)$ into two parts by the law of total variance to get
\bal
\xi_{d,\,k}^2(t_0)=\text{var}\big[ \text{E} \big\lbrace h(t_0;\,S) \mid  Z_{(1:d)}\big\rbrace \big] = \underbrace{\text{var}\big\lbrace h(t_0;\,S)\big\rbrace}_{V^{(h)}(t_0)} - \underbrace{\text{E}\big[\text{var}\big\lbrace h(t_0;\,S) \mid  Z_{(1:d)}\big\rbrace \big]}_{ \tilde{\xi}_{d,\,k}^2(t_0) }.\nonumber
\eal
Hence, the Hoeffding decomposition becomes
\bal
\text{var}\big\lbrace U(t_0) \big\rbrace
&= \sum_{d = 1}^k \binom{n}{k}^{-1} \binom{k}{d} \binom{n-k}{k-d} \,\,\xi_{d,\,k}^2(t_0)
= \sum_{d = 1}^k \gamma_{d,k,n} \left\lbrace V^{(h)}(t_0) - \tilde{\xi}_{d,\,k}^2(t_0) \right\rbrace,\nonumber
\eal
where \[\gamma_{d,k,n}=\binom{n}{k}^{-1} \binom{k}{d} \binom{n-k}{k-d}\] corresponds to the probability mass of a hyper-geometric distribution with parameters $n$, $k$, and $d$.
Denote \[V^{(S)}(t_0) = \sum_{d = 0}^k \gamma_{d,k,n} \,\,\tilde{\xi}_{d,\,k}^2(t_0),\]
then, after incorporating the $d=0$ term, we have
\bal
\text{var}\big\lbrace U(t_0) \big\rbrace
&=\left(1-\gamma_{0,k,n}\right) V^{(h)}(t_0) - \left\lbrace V^{(S)}(t_0)-\gamma_{0,k,n} \,\,\tilde{\xi}_{0,\,k}^2(t_0)\right\rbrace \nonumber\\
&= V^{(h)}(t_0) - V^{(S)}(t_0).\nonumber
\eal

This result is suitable for point estimation of a survival probability but does not have proper coverage of a confidence band, in which the variance estimation becomes a U-process. To construct a confidence band for a U-process $U(\mathbf{t})=\lbrace U(t_p):p=1,\ldots,P \rbrace$, we need to estimate the covariance of U-statistics at any two points, i.e., $\text{cov}\lbrace U(t_1),\,U(t_2)\rbrace$.
Given $t_1\neq t_2$, the covariance of $U(t_ 1)$ and $U(t_ 2)$ can be decomposed similarly to the above derivation (see appendix for details). Denote $C^{(h)}(t_1,\,t_2)=\text{cov}\lbrace h(t_1;\,S),\,h(t_2;\,S)\rbrace$, and
\[C^{(S)}(t_1,\,t_2)=\sum_{d = 0}^k \gamma_{d,k,n} \,\text{E}\big[\text{cov}\big( h(t_1;\,S),\,h(t_2;\,S) \mid  Z_{(1:d)}\big) \big]
=\sum_{d = 0}^k \gamma_{d,k,n} \,\,\tilde{\xi}_{d,\,k}^2(t_1,\,t_2),\] we have
\begin{align}\label{eq:cov_t1t2}
\text{cov}\big\lbrace U(t_1),\,U(t_2)\big\rbrace =C^{(h)}(t_1,\,t_2)-C^{(S)}(t_1,\,t_2).
\end{align}
When $t_1=t_2$, this equation reduces to the variance formula. The remaining challenge is to estimate both $C^{(h)}(t_1,\,t_2)$ and $C^{(S)}(t_1,\,t_2)$. To deal with the first term, let's consider a pair of  $S_i$ and $S_j$ that do not contain any overlapping samples. Then
\begin{equation}
C^{(h)}(t_1,\,t_2)
=\text{E}\left[\frac{1}{2}\Big\lbrace h(t_1;\,S_i)-h(t_1;\,S_j)\Big\rbrace\Big\lbrace h(t_2;\,S_i)-h(t_2;\,S_j)\Big\rbrace\right].\label{eq:Ch}
\end{equation}

Hence, we can estimate $C^{(h)}(t_1,\,t_2)$ using all disjoint pairs of subsamples. $C^{(S)}(t_1,\,t_2)$, on the other hand, involves subsamples with overlaps. For a pair of subsamples $S_i$ and $S_j$ such that $S_i\cap S_j=Z_{(1:d)}$,
\bal
\tilde{\xi}_{d,\,k}^2(t_1,\,t_2)&=\text{E}\big[\text{cov}\big\lbrace h(t_1;\,S),\,h(t_2;\,S) \mid  Z_{(1:d)}\big\rbrace \big]\nonumber\\
&=\text{E}\left[\frac{1}{2}\Big\lbrace h(t_1;\,S_i)-h(t_1;\,S_j)\Big\rbrace \Big\lbrace h(t_2;\,S_i)-h(t_2;\,S_j)\Big\rbrace\right]\nonumber.
\eal
We can use all such pairs of subsamples to estimate $\tilde{\xi}_{d,\,k}^2(t_1,\,t_2)$ as
\[\hat{\tilde{\xi}}_{d,\,k}^2(t_1,\,t_2)=\frac{1}{N_d}\sum_{\mid S_i\cap S_j\mid =d}\frac{1}{2}\Big\lbrace h(t_1;\,S_i)-h(t_1;\,S_j)\Big\rbrace\Big\lbrace h(t_2;\,S_i)-h(t_2;\,S_j)\Big\rbrace,\]
where \[N_d=\binom{n}{k}\binom{k}{d} \binom{n-k}{k-d}.\]
This formula is only for a fixed $d$. In the Hoeffding decomposition, $d$ ranges from 1 to $k$. However, estimating all $d$ terms is not needed. In fact, by reorganizing the terms, we realize that $C^{(S)}(t_1,\,t_2)$ can be estimated by collecting all size $k$ subsamples. Hence, there is no need to differentiate them with the overlapping size $d$. The following theorem shows how to obtain the estimator of the covariance matrix. The proof is in the Appendix.

\begin{theorem}
\label{thm:covforU}
For a complete U-statistic $U(t)$ as defined in \eqref{eq:Un}, given fixed $t_1\neq t_2$, the covariance of $U(t_1)$ and $U(t_2)$ can be estimated as
\[\hat{\text{cov}}\big\lbrace U(t_1),\,U(t_2)\big\rbrace=\hat{C}^{(h)}(t_1,\,t_2)-\hat{C}^{(S)}(t_1,\,t_2),\]
and
\begin{align*}
\hat{C}^{(h)}(t_1,\,t_2)
 &= \hat{C}^{(h)}(t_1,\,t_2)=\frac{1}{2N_0}\sum_{S_i\cap S_j=\emptyset}
\Big\lbrace h(t_1;\,S_i)-h(t_1;\,S_j)\Big\rbrace \Big\lbrace h(t_2;\,S_i)-h(t_2;\,S_j)\Big\rbrace,\\
\hat{C}^{(S)}(t_1,\,t_2)
 &= \frac{1}{N}\sum_{S_i\subset\cD_n}\Big\lbrace h(t_1;\,S_i)-U(t_1)\Big\rbrace\Big\lbrace h(t_2;\,S_i)-U(t_2)\Big\rbrace,
\end{align*}
where
\[N_0=\binom{n}{k} \binom{n-k}{k},\, N=\binom{n}{k}.\]
\end{theorem}

From Theorem \ref{thm:covforU}, both $\hat{C}^{(h)}(t_1,\,t_2)$ and $\hat{C}^{(S)}(t_1,\,t_2)$ can be estimated by re-sampling from their respective collection of subsamples. However, we need to restrict $k$ to $k\leq n/2$ since the estimation of $\hat{C}^{(h)}(t_1,\,t_2)$ requires independent subsets.

\subsection{Random Forests as Incomplete U-Statistics}

Similarly, we can develop an unbiased incomplete estimator without exhausting all subsamples. A random forest model usually fits a pre-specified large number of trees $B$. Denote the incomplete U-statistic as $U_B$. We can use the matched disjoint sampling scheme from \cite{xu2022variance} to obtain $B$ sets of independent subsamples. \cite{wang2014variance} also used this sampling scheme, and it can even be traced back to \cite{folsom1984probability} in the sampling design literature. The essential idea is to sample pairs of disjoint subsets, and use them to estimate the tree variance part $\hat{C}^{(h)}(t_1,\,t_2)$, while the samples cross different pairs can have overlaps and will be used to estimate the sample variance term $\hat{C}^{(S)}(t_1,\,t_2)$. The sampling procedure is outlined in algorithm \ref{alg:samp}.

\begin{algorithm}
\caption{Sampling scheme for obtaining pairwise matched disjoint subsamples.}\label{alg:samp}
\begin{tabbing}
\qquad \enspace \textbf{Input}: number of trees $B$, training data $\cD_n$, and kernel order $k$\\
   \qquad \enspace For $b=1$ to $b=B$\\
      \qquad \qquad Draw subsample $S_{b1}\subset \cD_n$ of size $k$\\
      \qquad \qquad Draw subsample $S_{b2}\subset \cD_n\setminus S_{b1}$ of size $k$. \\
\qquad \enspace \textbf{Output}: $\{S_{b1},\,S_{b2}\}_{b=1}^B$
\end{tabbing}
\end{algorithm}

Based on these incomplete matched subsamples $\{S_{b1},\,S_{b2}\}_{b=1}^B$, we define an incomplete U-statistic as
\[U_{B}(t)=\frac{1}{B}\sum_{b=1}^B\frac{1}{2}\Big\lbrace h(t;\,S_{b1})+h(t;\,S_{b2})\Big\rbrace.\]
The difference between the variance of an incomplete U-statistic and its complete counterpart \citep{Lee1990U-statistics:Practice} is
\[\text{var}(U_{B})=\text{var}\Big\lbrace \text{E}\,(U_{B}\mid \cD_n)\Big\rbrace+\text{E}\Big\lbrace\text{var}(U_{B}\mid \cD_n)\Big\rbrace=\text{var}(U)+\text{E}\Big\lbrace\text{var}(U_{B}\mid \cD_n)\Big\rbrace.\]
Hence for any $t_1\neq t_2$,
\begin{eqnarray*}
\text{cov}\big\lbrace U_{B}(t_1),\,U_{B}(t_2)\big\rbrace
&=&\text{cov}\Big[\text{E}\,\big\lbrace U_{B}(t_1)\mid \cD_n\big\rbrace,\,\text{E}\big\lbrace U_{B}(t_2)\mid \cD_n\big\rbrace\Big]+\\
&&\text{E}\Big[\text{cov}\big\lbrace U_{B}(t_1),\,U_{B}(t_2)\mid \cD_n\big\rbrace\Big]\\
&=&\text{cov}\big\lbrace U(t_1),\,U(t_2)\big\rbrace + \text{E}\Big[\text{cov}\big\lbrace U_{B}(t_1),\,U_B(t_2)\mid \cD_n\big\rbrace\Big].
\end{eqnarray*}
The last term in the above equation is the bias, which we need to estimate. The following theorem shows the calculation of the bias. The proof is in the appendix.

\begin{theorem}\label{thm:biasUB}
For the incomplete U-statistic $U_B(t)$ defined on the pairwise matched group subsamples, given $t_1$ and $t_2$,
\begin{align*}
\text{E}\Big[\text{cov}\big\lbrace U_{B}(t_1),\,U_B(t_2)\mid \cD_n\big\rbrace \Big]
=\frac{1}{2B}\,C^{(h)}(t_1,\,t_2)-\frac{1}{B}\text{cov}\,\big\lbrace U(t_1),\,U(t_2)\big\rbrace.
\end{align*}
Furthermore,
\begin{align*}
\text{cov}\big\lbrace U_{B}(t_1),\,U_B(t_2)\big\rbrace = \left(1-\frac{1}{B}\right)\text{cov}\,\big\lbrace U(t_1),\,U(t_2)\big\rbrace +
\frac{1}{2B}\,C^{(h)}(t_1,\,t_2).
\end{align*}
\end{theorem}

The estimators $\hat{C}^{(h)}(t_1,\,t_2)$ and $\hat{C}^{(S)}(t_1,\,t_2)$ in Theorem \ref{thm:covforU} are constructed based on all subsamples of sample size $k$ in the data set $\cD_n$. However, the matched disjoint sampling scheme only yields incomplete subsamples. Therefore, we need to reconstruct new estimators for $C^{(h)}(t_1,\,t_2)$ and $C^{(S)}(t_1,\,t_2)$ based only on these incomplete subsamples $\{S_{b1},\,S_{b2}\}_{b=1}^B$.

From the equation (\ref{eq:Ch}),
an unbiased estimator of $C^{(h)}(t_p)$ can be defined as
\begin{align}\label{eq:chb}
\hat{C}^{(h)}_{B}(t_1,\,t_2)=&\frac{1}{2B}\sum_{b=1}^B\Big\lbrace h(t_1;\,S_{b1})-h(t_1;\,S_{b2})\Big\rbrace \Big\lbrace h(t_2;\,S_{b1})-h(t_2;\,S_{b2})\Big\rbrace.
\end{align}
For $C^{(S)}(t_1,\,t_2)$, we define
\begin{align}\label{eq:csb}
\hat{C}^{(S)}_{B}(t_1,\,t_2)=&\frac{1}{2B}\sum_{b=1}^B\sum_{i=1}^2\Big\lbrace h(t_1;\,S_{bi})-U_{B}(t_1)\Big\rbrace\Big\lbrace h(t_2;\,S_{bi})-U_{B}(t_2)\Big\rbrace,
\end{align}
which is an biased estimator since the subsamples $S_{b1}$ and $S_{b2}$ are dependent.
The following theorem further quantifies this bias so that we can correct it. The proof is in the Appendix.

\begin{theorem}\label{thm:biasVS}
For the sample variance estimator $\hat{C}^{(S)}_{B}(t_1,\,t_2)$ defined on the pairwise matched disjoint subsamples, we have
\begin{align*}
\text{E}\Big\lbrace\hat{C}^{(S)}_{B}(t_1,\,t_2)\Big\rbrace=\left(1-\frac{1}{B}\right)C^{(S)}(t_1,\,t_2) + \frac{1}{2B}\,C^{(h)}(t_1,\,t_2).
\end{align*}
\end{theorem}

\vskip 0.3cm

From Theorem \ref{thm:biasUB} and equation (\ref{eq:cov_t1t2}), we obtain
\begin{eqnarray*}
\text{cov}\big\lbrace U_{B}(t_1),\,U_B(t_2)\big\rbrace
&=& \left(1-\frac{1}{B}\right)\text{cov}\big\lbrace U(t_1),\,U(t_2)\big\rbrace +
\frac{1}{2B}\,C^{(h)}(t_1,\,t_2)\\
&=& \left(1-\frac{1}{B}\right)\left\lbrace C^{(h)}(t_1,\,t_2) -   C^{(S)}(t_1,\,t_2)\right\rbrace + \frac{1}{2B}\,C^{(h)}(t_1,\,t_2)\\
&=& C^{(h)}(t_1,\,t_2) - \left(1-\frac{1}{B}\right)  C^{(S)}(t_1,\,t_2) - \frac{1}{2B}\,C^{(h)}(t_1,\,t_2).
\end{eqnarray*}
Together with Theorem \ref{thm:biasVS}, we get an unbiased estimator of $\text{cov}\lbrace U_{B}(t_1),\,U_B(t_2)\rbrace$,
\begin{align*}
\hat{\text{cov}}\lbrace U_{B}(t_1),\,U_B(t_2)\rbrace
= \hat{C}_B^{(h)}(t_1,\,t_2) - \hat{C}_B^{(S)}(t_1,\,t_2).
\end{align*}

\section{Estimating the Critical Value}

Obtaining the critical value for a confidence band is a challenging issue. \cite{Koul2006FittingModels} uses a Gaussian process whose covariance function depends on a fitted model and fits confidence bands with a time series model. \cite{Hall1980ConfidenceData} applies the weak convergence of the Kaplan-Meier survival estimator to a Gaussian process with a specific covariance functional to form simultaneous confidence bands. Many other papers also use Gaussian processes as asymptotic convergence of survival functions and obtain confidence bands \citep{Bose2002, Zhu2002ResamplingModel, Conde-Amboage2021AData}. However, the calculations of these confidence bands are not straightforward, particularly in \cite{Koul2006FittingModels}. An alternative approach that is promising in our setting is proposed in \cite{lin1994confidence}, which numerically generates samples to approximate the Gaussian Process under the Cox model \cite{cox1972regression}.

Apart from the computational complexity, these methodologies do not work directly for our current setting since no specific covariance matrix structure is assumed. We may reasonably expect $U_B(t)$ to converge weakly to a Gaussian process. Note that this would still be a strong statement, given that there is a lack of literature on the asymptotic results of random survival forests \citep{steingrimsson2019censoring, Cui2019}. Furthermore, the theoretical properties of $U_B$  under $k = O(n)$ is not well understood even in the regression setting. Hence, it is difficult to evaluate the limiting distribution analytically and obtain the critical value. Therefore, we propose a simulated approach based on the estimated covariance matrix to obtain a critical value in the confidence band estimation empirically.

The procedure works as follows. First, we get an estimator of the covariance matrix of $U_B(t)$ at $P$ time points, $t_1,\,\ldots,\,t_P$.
Next, for a significance level $\alpha$, we need to find a $\zeta_{\alpha}$ that satisfies
\begin{align*}
\text{pr}\left\lbrace\left|\frac{U_B(t_p)-\text{E}\,\lbrace\,U_B(t_p)\,\rbrace}{[\,\text{var}\lbrace U_B(t_p)\rbrace\,]^{1/2}}\right| \leq \zeta_{\alpha},\,\,p=1,\ldots,P\right\rbrace\geq 1-\alpha.
\end{align*}
Then the confidence band of predicted function $U_B(t)$ under the significance level $\alpha$ is
\begin{align*}
\left\{U_B(t_p):\,\,\text{E}\,\lbrace\,U_B(t_p)\,\rbrace \pm \zeta_{\alpha}[\,\text{var}\lbrace U_B(t_p)\rbrace\,]^{1/2},\,\, p=1,\ldots,P\right\}.
\end{align*}

We first outline the covariance matrix estimation, which is given in Algorithm \ref{alg:cov}.

\begin{algorithm}
\caption{The algorithm for calculating the predicted curve $U_B(t)$ and the covariance matrix estimator $\hat\Sigma$.}\label{alg:cov}
\begin{tabbing}
   \qquad \enspace \textbf{Input}: training data $\cD_n$, kernel order $k$, and prediction time points $t_1,\,\ldots,\,t_P$\\
   \qquad \enspace For $b=1$ to $b=B$ \\
   \qquad \qquad Sample $S_{b1}$ and $S_{b2}$ according to algorithm \ref{alg:samp}\\
   \qquad \qquad Fit random trees $h(\cdot,\,S_{b1})$ and $h(\cdot,\,S_{b2})$ \\
   \qquad \qquad Calculate $h(t,S_{bi})=(h(t_1,\,S_{bi}),\ldots,h(t_P,\,S_{bi}))$\\
   \qquad \qquad Calculate the average of $h(t, S_{bi})$ to get  $\bar{h}_b = \big\lbrace h(t, S_{b1})+h(t, S_{b2})\big\rbrace/2$\\
\qquad \enspace Calculate the random forest estimator \\
\qquad \qquad $ U_{B}(t)=B^{-1}\sum_{b=1}^B\bar{h}_b$ \\
\qquad \enspace Calculate the covariance matrix for $1 \leq j \leq l \leq P$\\
\qquad \qquad $\hat{C}^{(h)}_{B}(t_j,\,t_l)$ using Equation \eqref{eq:chb}\\
\qquad \qquad $\hat{C}^{(S)}_{B}(t_j,\,t_l)$ using Equation \eqref{eq:csb}\\
\qquad \qquad $\hat\Sigma_{jl} = \hat\Sigma_{lj} = \hat{C}^{(h)}_{B}(t_j,\,t_l) - \hat{C}^{(S)}_{B}(t_j,\,t_l)$\\

\qquad \enspace \textbf{Output}: the estimator $U_{B}(t)$, and the covariance matrix $\hat\Sigma$
\end{tabbing}
\end{algorithm}

Two difficulties remain. First, $\hat\Sigma$ may not be positive definite due to the numerical estimation. As such, we can project $\hat\Sigma$ to the nearest positive definite matrix $\hat\Sigma^{(+)}$. This transformation will also force positive estimates for the marginal variances along the diagonal and allow us to use the covariance matrix for sampling from a multivariate normal distribution. Secondly, the variance estimation of two adjacent time points could change rapidly due to numerical instabilities. In some examples in the simulation study, the variance estimation is close to 0, causing sudden low coverage at specific time points. This issue could be alleviated by imposing smoothness over the time domain. To be precise, we perform kernel smoothing on the diagonal of the matrix $\hat\Sigma^{(+)}$ with the optimal bandwidth obtained by \citep{Silverman1986}. The updated covariance matrix with the smoothed diagonal is denoted as $\hat\Sigma^{(+s)}$. These two covariance matrix estimators will be used for estimating the critical value in a confidence band. In our simulation study, We will evaluate the performance of both of them.

Next, we will use a sampling distribution to select the critical value $\zeta_{\alpha}$. For a sufficiently large number $M$, sample $W_1,\,\ldots,\, W_M$ from a multivariate normal distribution with mean vector $U_B(t)$ and covariance matrix $\hat\Sigma^{(+s)}$ (or $\hat\Sigma^{(+)}$). Denote the $p$th element of $W_i$ as $W_{ip}$, and the diagonal elements of the matrix $\hat\Sigma^{(+s)}$ as $(s_1^2,\,\ldots,\,s_P^2)$. For any given number $c$, we can calculate
\begin{align*}
M(c)= \text{Number of elements in set }
\left\{W_i:
\left| \frac{W_{ip}-U_B(t_p)}{s_k}\right| \leq c,\,\,  p=1,\ldots,\,P.\right\}.
\end{align*}
Then $M(c)/M$ measures the proportion of sampled curves enclosed entirely by the confidence band
\begin{align*}
\left\{U_B(t):\,\,U_B(t_p)\,\pm c\,s_k,\,\,  p=1,\ldots,\,P\right\},
\end{align*}
Hence, given the confidence level $\alpha$, the critical value $\zeta_\alpha$ can be obtained by
\bal
\zeta_\alpha = \inf_{0<c<\infty}\left\lbrace c:\, M(c)\geq (1-\alpha)M \right\rbrace\label{eq:critchoice}.
\eal
Then the confidence band at time point $t_p$ is
\begin{equation}
U_B(t_p)\pm \zeta_\alpha s_k.
\label{eq:band}
\end{equation}
This process is summarized in algorithm \ref{alg:conband}.

\begin{algorithm}
\caption{The algorithm for calculating the confidence band.}\label{alg:conband}
\begin{tabbing}
    \qquad \enspace \textbf{Input}: simulated sample size $M$, the lower and upper points $c_1$ and $c_2$,\\
    \qquad\,\, the estimates $U_B(t)$ and $\hat\Sigma$ calculated by Algorithm \ref{alg:cov}\\
    \qquad \enspace Project $\hat\Sigma$ to the nearest positive definite matrix $\hat\Sigma^{(+)}$ \\
    \qquad \enspace Perform Gaussian kernel smoothing on the diagonal of $\hat\Sigma^{(+)}$ to get $\hat\Sigma^{(+s)}$\\
    \qquad \enspace If $\hat\Sigma^{(+s)}$ is not positive definite\\
    \qquad \qquad Update $\hat\Sigma^{(+s)}$ by projecting to the nearest positive definite matrix again\\
    \qquad \enspace For $i=1$ to $i=M$ \\
    \qquad \qquad Draw $V_i \sim \text{MVN}(U_B(t), \hat\Sigma^{(+s)})$\\
    \qquad \enspace For $c=c_1$ to $c=c_2$ \\
    \qquad \qquad Calculate $M(c)$\\
    \qquad \enspace Choose $\zeta_\alpha$ according to \eqref{eq:critchoice}\\
    \qquad \enspace \textbf{Output}: Confidence band \eqref{eq:band}
\end{tabbing}
\end{algorithm}

\section{Numerical Studies}

\subsection{Simulation Settings}\label{sec:setup}

There are no existing methods to provide a random survival forest confidence band. Hence our main goal here is to evaluate the coverage rate and understand the performance of the proposed method under different tuning parameters. For all simulation settings, we use $n=1000$ independent samples. We generate six covariates independently from a Uniform$(0,\,1)$ distribution for each observation. We then generate the failure time $T$ from one of the distributions summarized in Table \ref{tab:scenarios}. The censoring time $C$ is generated from an exponential distribution with parameter $\lambda$. Then we obtain the observed survival time $Y=\min(T,\,C,\,\tau)$, and censoring indicator $\delta=I(T\leq C)$, where $\tau$ is set as a fixed value to control the maximum follow-up time \citep{cui2020estimating}. These specifications and the resulting failure rates are also summarized in Table \ref{tab:scenarios}.

\begin{table}[ht]
\caption{Simulation Settings}
\label{tab:scenarios}
\begin{tabular}{cllccc}
Scenario & Distribution & Parameters                                       & $\tau$ & $\lambda$ & Failure Rate \\
1        & Gamma        & rate 1, shape $2x^{(1)}$                       & 2.50    & 0.25      & 0.76\\
2        & Log-Normal   & mean $e$, std. dev. $e^{x^{(1)}}$     & 7.50    & 0.10       & 0.73\\
3        & Log-Normal   & mean $e^{x^{(1)}-0.5}$, std. dev.  $e$ & 7.50    & 0.10       & 0.85\\
4        & Weibull      & scale 1, shape $5x^{(1)}$                      & 2.00      & 0.25      & 0.77\\
\end{tabular}
\end{table}

We apply the confidence band calculation on three targets points and their cumulative hazard functions: $x_1=(0.25,\,0.50,\,\ldots,\,0.50)^T$, $x_2=(0.50,\,0.50,\,\ldots,\,0.50)^T$ and $x_3=(0.75,\,0.50,\,\ldots,\,0.50)^T$. Except for the first coordinate $x^{(1)}$, none of the other variables are important. Hence, we set all other predictor variables to 0.5. In each scenario, we set $k=500$ and $B=10,000$ as the number of pairwise matched subsamples, and $M=1,000$ for the size of multivariate normal samples. For the random forest settings, we use the best split strategy from \citep{ishwaran2008random}. We also used three variables to be considered at each split and a minimum node size of 15.

Our method is implemented using the R package \texttt{RLT}, available on GitHub at \url{https://github.com/teazrq/RLT}. Each simulation is repeated $R=1,000$ times for each scenario. To evaluate the performance, we calculate the coverage probabilities of the confidence band, evaluated on a grid of time points $t=(t_1,\,\ldots,\,t_{500})=(0,\,\tau/500,\,\ldots,\,499\tau/500)$. The coverage probability is estimated by
\begin{equation}
    \frac{1}{R}\sum_{i=1}^{R}I\left\{ \Lambda_0(t_p)\in\left(\Lambda_L^{(r)}(t_p),\,\Lambda_U^{(r)}(t_p)\right),\,\, p=1,\,\ldots,\,500.\right\},\nonumber
\end{equation}
where $\Lambda_0(t_p)$ is the random forest mean cumulative hazard at $t_p$, and $\left(\Lambda_L^{(r)}(\cdot),\,\Lambda_U^{(r)}(\cdot)\right)$ is the confidence band obtained at the $r$th iteration. In the case of our simulations, we have the true distribution for each scenario. However, random forests estimators can be biased \citep{Cui2019}. Given a set of tuning parameters, we do not know the bias theoretically. Hence, we instead approximate the truth $\Lambda_0(t_p)$ using the empirical mean of all $R=1000$ random forest estimators. This empirical mean is indeed very close to the true cumulative hazard function, as we can see from Fig. \ref{fig:avg_true} in the Appendix. The bias is most prominent in the tails, so we would expect the coverage of the truth in earlier time points to be similar to the coverage of the expected random forest. Since potential bias can comprise the validity of the covariate rate estimation, we use this approximation of the mean random forest estimator as the true mean value of the mean cumulative hazard. In the following, we analyze the performance from three aspects: the overall coverage rate, the variance of the proposed estimator, and the effect of the number of trees.

\subsection{Coverage Rates}
Table \ref{tab:coverage} summarizes the coverage probabilities for the $95\%$ confidence bands obtained by two covariance matrix correction methods under different scenarios. Original projection means only projecting $\hat\Sigma$ to the nearest positive definite matrix $\hat\Sigma^{(+)}$ without smoothing correction, while smoothed projection means combining projection and smoothing corrections. We can see that the original projection method has slightly lower coverage probabilities than the target of $95\%$. In contrast, the confidence bands constructed by the smoothed projection method included the random forest truth in at least $95\%$ of the simulations for every scenario. These results show the power of our proposed methodology for producing functioning confidence bands.

\begin{table}[ht]
\caption{The coverage probabilities for the $95\%$ confidence bands of the cumulative risk functions at different values of $x_i=(x_i^{(1)},\,0.50,\,\ldots,\,0.50)^T$ under different scenarios.}\label{tab:coverage}
\centering
\begin{tabular}[t]{cccc}
  & $x_i^{(1)}$ & Original Projection & Smoothed Projection\\
Scenario 1 & 0.25 & 0.913 & 0.954\\
 & 0.50 & 0.947 & 0.977\\
 & 0.75 & 0.944 & 0.976\\
Scenario 2 & 0.25 & 0.884 & 0.971\\
 & 0.50 & 0.898 & 0.981\\
 & 0.75 & 0.869 & 0.981\\
Scenario 3 & 0.25 & 0.901 & 0.967\\
 & 0.50 & 0.921 & 0.981\\
 & 0.75 & 0.871 & 0.968\\
Scenario 4 & 0.25 & 0.940 & 0.972\\
 & 0.50 & 0.943 & 0.981\\
 & 0.75 & 0.930 & 0.974\\
\end{tabular}
\end{table}

\begin{figure}[ht]
    \centering
    \includegraphics{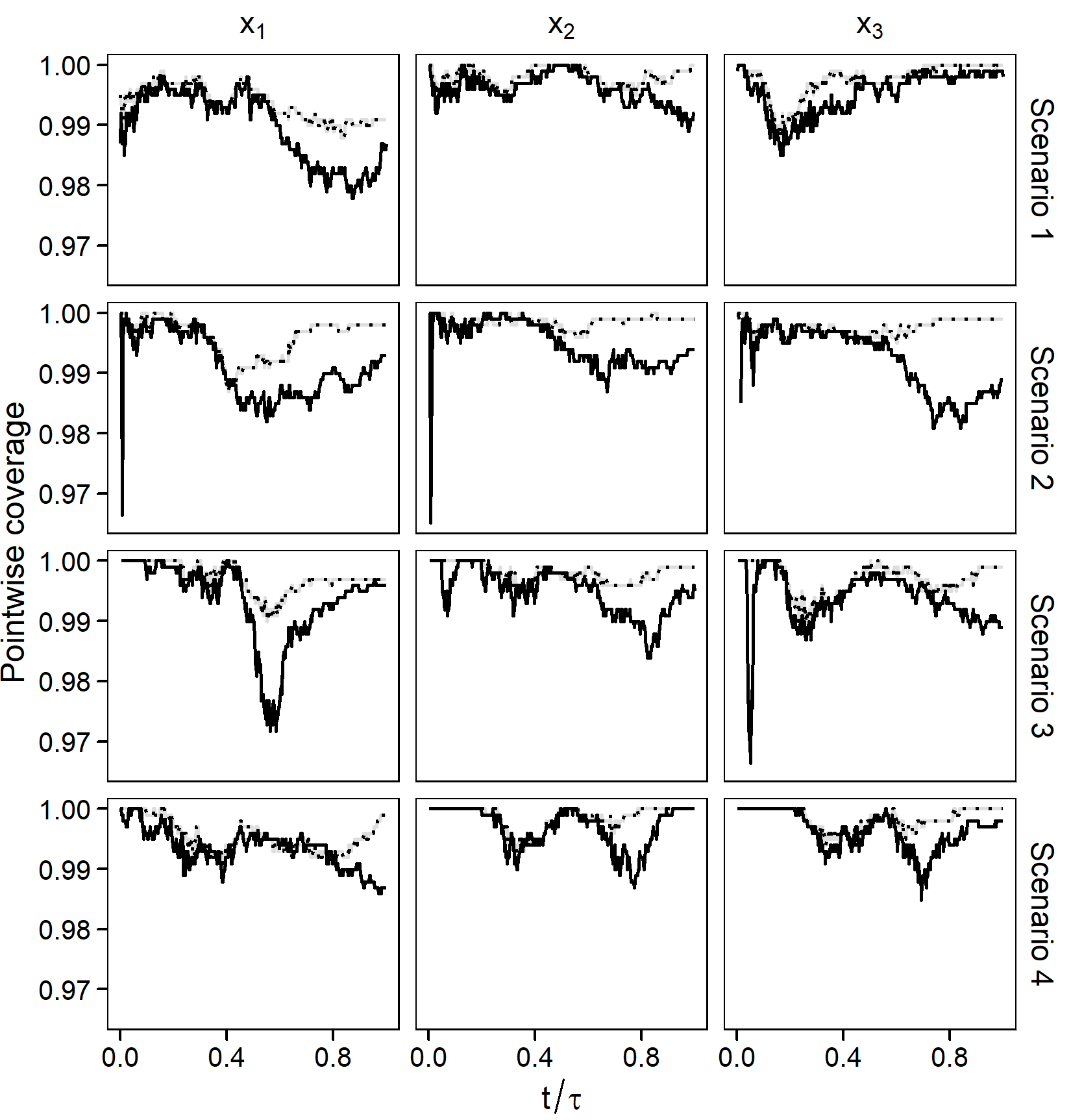}
    \caption{Pointwise coverage probabilities obtained by the original projection method (solid) and the smoothed projection method (dots) over time.}
    \label{fig:pw_cover}
\end{figure}

To identify the time points that tend to fall out of the confidence bands, we provide Fig. \ref{fig:pw_cover}, which gives the pointwise coverage probabilities over time based on the original projection and the smoothed projection correction methods. A y-axis limit of $0.97\leq y\leq 1$ is used for a simplified presentation, which cuts off some time points in Scenarios 2 and 3, the log-Normal scenarios. Because the critical value we choose, $\zeta_\alpha$, is based on the entire cumulative risk curve and not a single point, the pointwise coverage probabilities are much higher than the nominal level $\alpha=95\%$. The lowest pointwise coverage is for Scenario 2, where the original projection covers only 0.94 at $t=0.01\tau$ when $x^{(1)}=0.75$.
The original projection method shows its weakness at many earlier time points with sharp drops in Scenarios 2 and 3. The smoothed projection method prevents abrupt changes, and the pointwise coverage probabilities are almost always better than that of the original projection method.
The weaker performance for the smoothed projection method is in the middle time points, with both tails having pointwise coverage probabilities over $99\%$.
However, the smoothed projection method has nearly perfect pointwise coverage probabilities in the later time points for some of the scenarios, one possible reason for which we will discuss in section \ref{sec:margvar}.

\subsection{Marginal Variance Analysis}\label{sec:margvar}

First, we present a randomly selected 25 confidence bands constructed by the smoothed projection method for each scenario to demonstrate their variations. These results are shown in Fig. \ref{fig:manyCB} in the Appendix. There are cases where the bands came close to excluding the random forest truth, but there are very few bounds that cross the random forest truth. Scenario 2 has the most simulations in this visual that are very close or fail to contain the truth, but it still achieves good coverage.

Furthermore, we compare two corrections of the covariance matrix: the original and smoothed projections. The primary problem with estimating the marginal variance is that some estimates are negative values. In our simulation, the proportion of negative marginal variances for these scenarios ranged from approximately $0.1\%$ (Scenario 1 when the target point is $x_1$) to 8.2\% (Scenario 2 when the target point is $x_1$).

\begin{figure}[ht]
    \centering
    \includegraphics{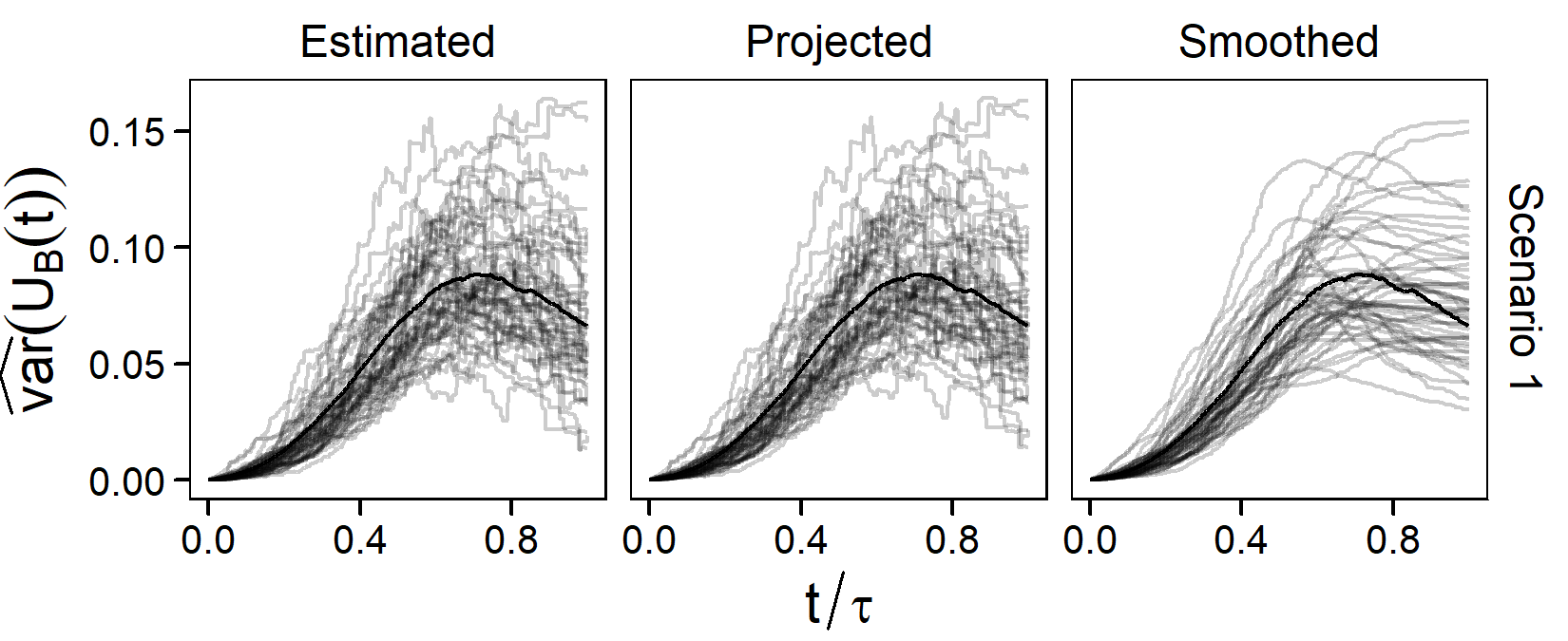}
    \caption{ Two-step correction procedure for the marginal variance of 25 simulations (grey) in Scenario 1. From left to right: original marginal variances, projected marginal variances, and final estimated marginal variances after projection and smoothing. True marginal variance in black. } 
    \label{fig:est_proj_smooth}
\end{figure}

Figure \ref{fig:est_proj_smooth} shows the process of the corrections. The left plot shows the original marginal variance estimators from 25 simulations in Scenario 1, some of which are negative. The middle plot gives the marginal variances from the same simulations after projection. The negative variance estimators in the left plot have been corrected. The marginal variances are visually almost identical except for the negative marginal variances. The right plot shows the estimation results combining projection and smoothing, which cleans up some of the sharp drops and spikes in the variance estimations.

Next, we examine how close the estimated marginal variances are to the actual marginal variance of random forest predictions. To find the actual variance of the random forest predictions, we calculate the marginal variance at time $t_p$ by calculating the sample variance of the random forest predictions at time $t_p$ across all simulations. Figure \ref{fig:var_avg} compares the sample variance curve overall simulations to the average of the estimated variances across all simulation runs. Except for $x_1$ in Scenario 4, the initially projected variance closely tracks the truth. The smoothed projected variance also closely follows the actual variance at earlier time points. However, the smoothed projected variance is biased at later time points and, in most cases, overestimates the true variance. This bias leads to over-coverage at later time points and the nearly 100\% coverage in Fig. \ref{fig:pw_cover}. Although smoothing introduces bias, it does bring the overall coverage probabilities in Table \ref{tab:coverage} to the desired level. Therefore, we still prefer the smoothed projected variance.

\subsection{Number of Trees and Negative Variance}

\begin{figure}
     \begin{subfigure}[t]{\textwidth}
         \centering
         \includegraphics{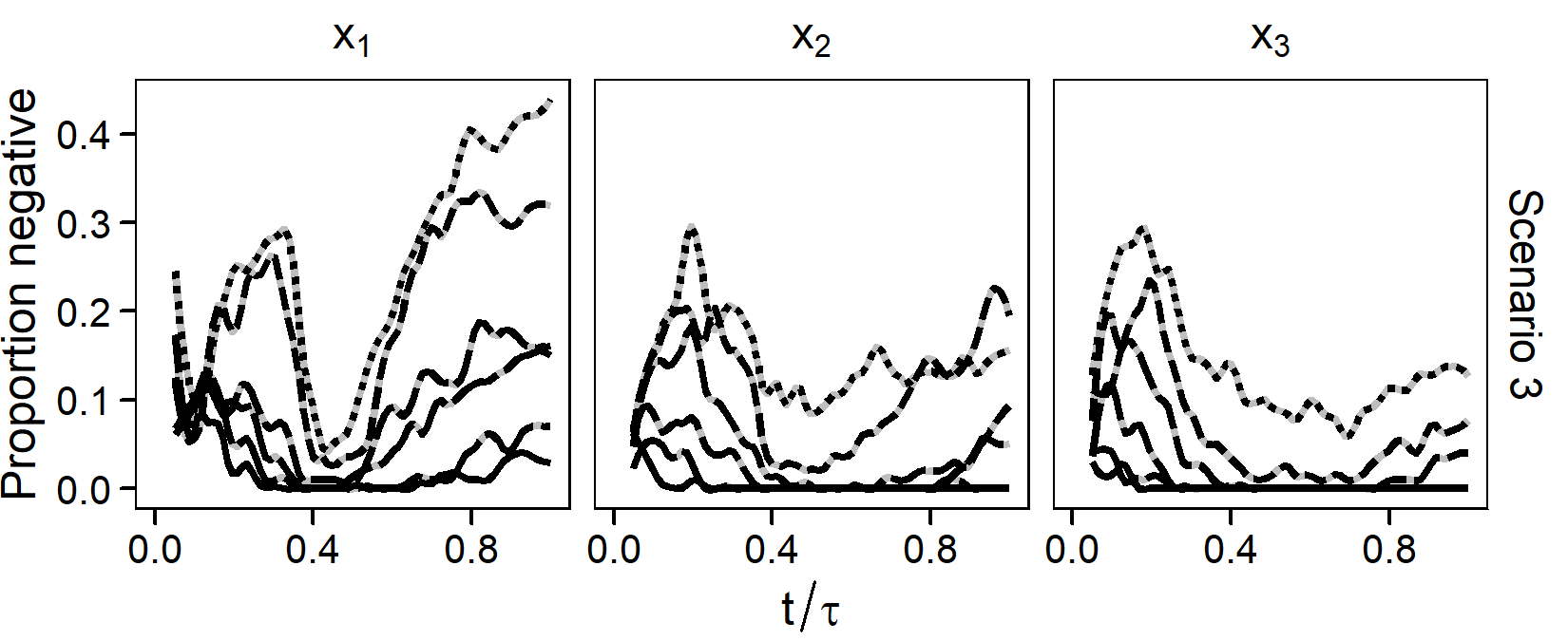}
         \caption{The pointwise proportions of negative estimated marginal variances.}
         \label{fig:neg_var_nt}
     \end{subfigure}
     \hfill
     \begin{subfigure}[t]{\textwidth}
         \centering
         \includegraphics{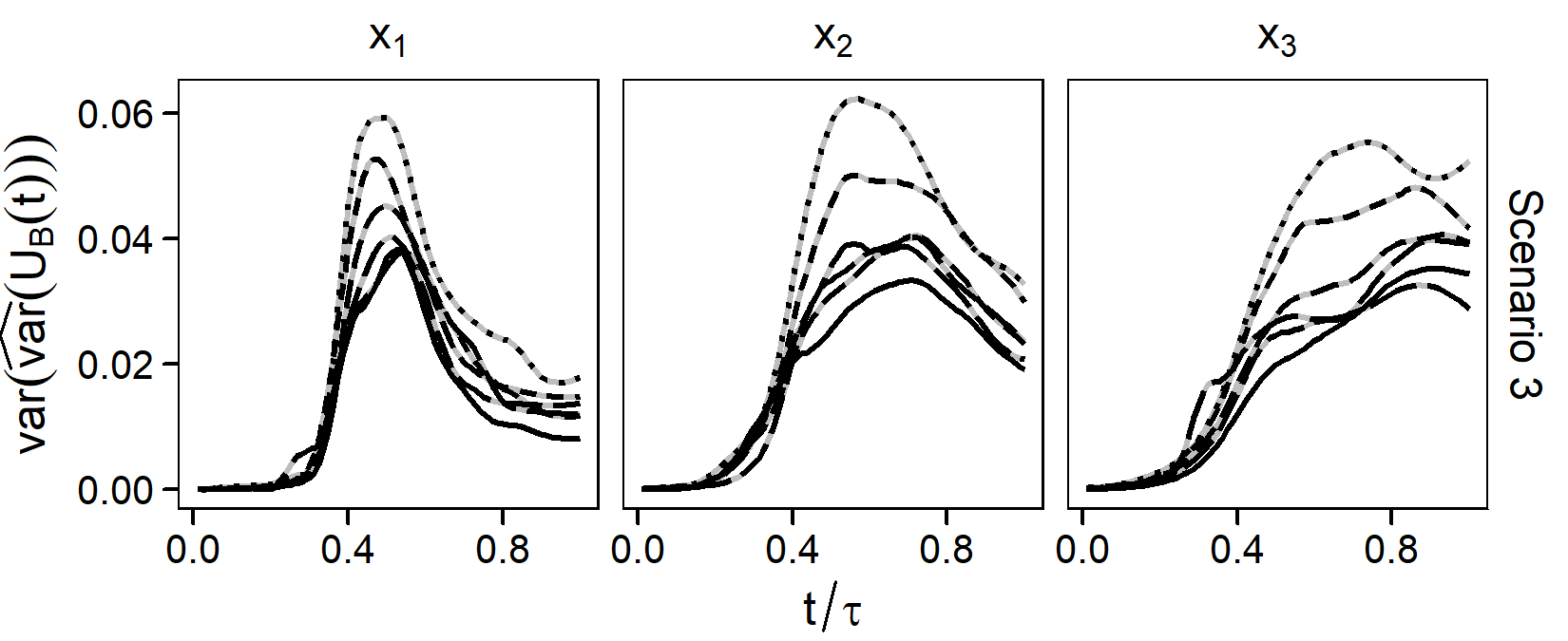}
         \caption{Variance of $\hat{\text{var}}(U_B(t))$ by number of trees.}
         \label{fig:var_avg_nt}
     \end{subfigure}
        \caption{Lines for $B=500,\,1,000,\,2,500,\,5,000,\,10,000$ and $20,000$, ascending in density of points with $B=500$ as single points and $B=20,000$ as a solid line.}
        \label{fig:twographs}
\end{figure}

Our final analysis shows the effect of the number of trees, $B$, for estimating the marginal variance. Focusing on Scenario 3, we re-ran the above analysis with several different values of $B$. The coverage probabilities of the estimated confidence bands
and the average proportions of negative estimated marginal variances in $\hat\Sigma$ before projection are listed in Table \ref{tab:neg_nt}, while the point-wise proportion is plotted in Fig. \ref{fig:neg_var_nt}. Fig. \ref{fig:var_avg_nt} shows the sample variance of the estimated marginal variance from $\hat\Sigma^{(+)}$.

\begin{table}
\caption{The coverage and proportions of negative estimated marginal variances under different numbers of trees.}\label{tab:neg_nt}
\centering
\begin{tabular}{ccccccc}
    &\multicolumn{3}{c}{Proportion of Negative Variance}&\multicolumn{3}{c}{Coverage Probability}\\
$B$ & $x^{(1)}$=0.25 & $x^{(1)}$=0.5 & $x^{(1)}$=0.75 & $x^{(1)}$=0.25 & $x^{(1)}$=0.5 & $x^{(1)}$=0.75\\
500 & 0.250 & 0.147 & 0.139 & 0.960 & 0.97 & 0.950\\
1,000 & 0.198 & 0.114 & 0.070 & 0.960 & 0.96 & 0.980\\
2,500 & 0.097 & 0.049 & 0.036 & 0.940 & 0.97 & 0.960\\
5,000 & 0.074 & 0.026 & 0.014 & 0.920 & 0.95 & 0.980\\
10,000 & 0.036 & 0.008 & 0.005 & 0.949 & 0.98 & 0.939\\
20,000 & 0.022 & 0.004 & 0.003 & 0.990 & 0.97 & 0.950\\
\end{tabular}
\end{table}

Firstly, as shown in Table \ref{tab:neg_nt} and Fig. \ref{fig:twographs}, a smaller number of trees leads to a higher proportion of negative estimated marginal variances. Both Fig. \ref{fig:neg_var_nt} and  Fig. \ref{fig:var_avg_nt} are loess smoothed with a span of 0.2 for clearer display of the trend. In Fig. \ref{fig:var_avg_nt}, the smaller the number of trees, the further the line is from 0, which implies that the smaller number of trees, the more variable the estimated marginal variance is. The highest proportion of negative variances and the most variable estimates are for $B=500$. This number is much smaller than generally recommended for random survival forests, so it is not surprising that it performs poorly on both metrics. There is a significant drop in the proportion of negative variances between $B=5,000$ and $B=10,000$. The setting we used for the simulations in Section \ref{sec:setup} is the second lightest line ($B=10,000$), with a low proportion of negative estimated marginal variances and a relatively stable estimator. Although the darkest line ($B=20,000$) shows the best result, the difference between these two in either metric is not much. Therefore, we used $B=10,000$ for the above simulations.

Even with high proportions of initial negative marginal variances, our methodology manages almost 95\% confidence without significant over-coverage caused by variance overestimation in the right tail.
\section{Real Data Analysis}\label{sec:realdata}
We further demonstrated our method using a data set from \cite{Kalbfleisch1980} on veteran's administration lung cancer data. This data is also used in \cite{ishwaran2008random}. The data is available in the R package \texttt{randomForestSRC} \citep{rfsrc} as \texttt{veteran}.
The response variable $y$ is days of survival. In the case of a subject dropping out of the study, $y$ is the days until drop-out. $n=137$ subjects had measurements for six predicted variables and a recorded death or last follow-up time. 6.6\% of the response times were censored.

We ran a random forest with the best split, a minimum terminal node size of 5, and 3 for the number of variables considered at each split. Based on this result, subjects 88 and 97 were selected as our test subjects to demonstrate their different predicted out-of-bag survival curves. We re-ran the random forest without those two subjects and calculated the estimated covariance matrices and confidence bands.

\begin{figure}
    \centering
    \includegraphics{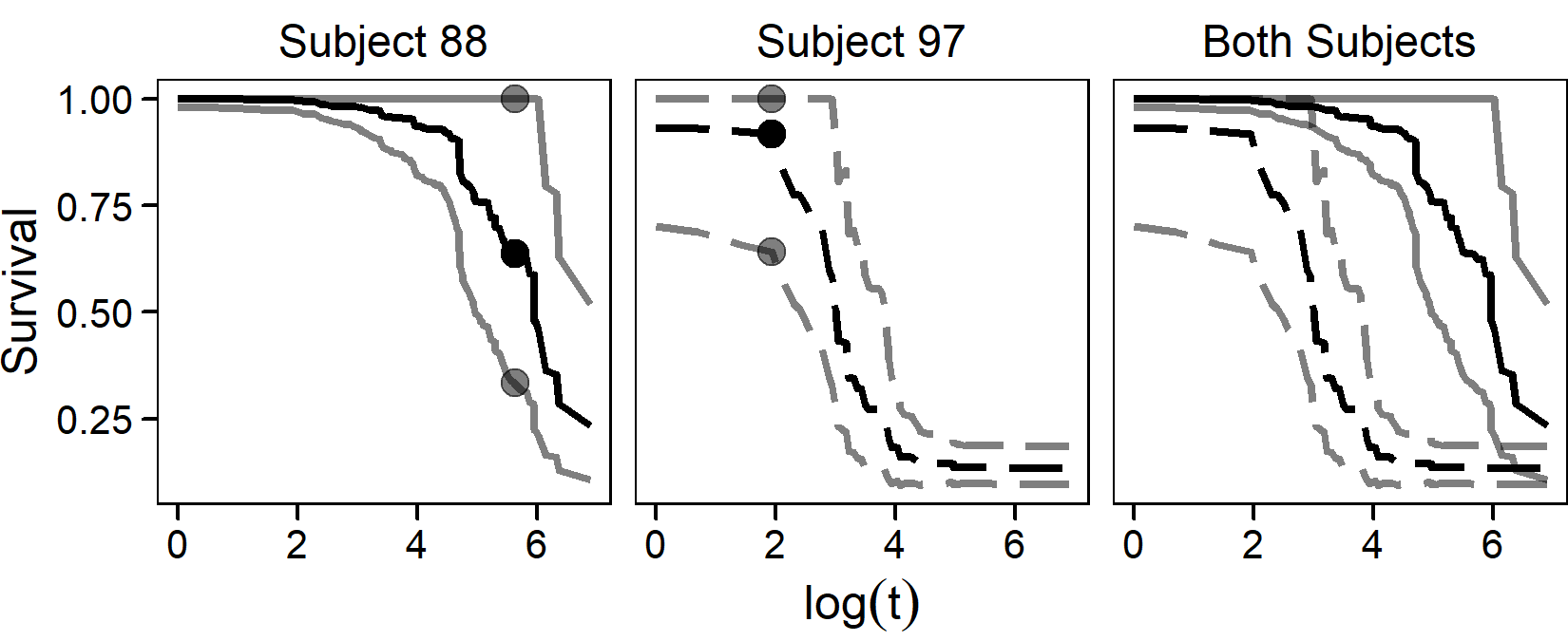}
    \caption{Estimated survival curves (black) and confidence bands (grey) for test subjects 97 (dots) and 88 (dashes). Actual survival time marked with a solid point.}
    \label{fig:real_subj}
\end{figure}

The confidence bands in Fig. \ref{fig:real_subj} are significantly different between the two subjects. Subject 88 was a 51-year-old with squamous lung cancer diagnosed two months before the study began and had a good Karnofsky performance score. His relatively long survival time of $y=283$ matches the slowly increasing cumulative hazard prediction and bands. The marginal variance keeps growing over time, showing a much wider range of possible cumulative hazard values at later time points. Subject 97, on the other hand, was a 66-year-old with small cell lung cancer diagnosed 11 months before the study and had a poor Karnofsky performance score. His short survival time of $y=7$ matches a much quicker increase in the cumulative hazard prediction and bands. The marginal variance stabilizes, and the prediction gives a consistent cumulative hazard range over time.

\section{Discussion}

We proposed a theoretically valid and unbiased confidence band estimator of random survival forests. This addresses the vacancy in the current literature. We also did not address the ratio consistency or rate of convergence of the covariance estimation since existing theory does not seem to provide enough tools or understanding of the $k=O(n)$ scenario. These theoretical considerations should be analyzed in the future.

Our method can be extended in several different directions. For example, the variable importance measure of a random survival forest is often done by calculating the C-index. However, these measures do not have appropriate associated confidence intervals, making inference difficult. As another direction, \cite{xu2022variance} considers $k>n/2$ for regression random forests using bootstrap approximations of the true variance. This can be incorporated into our approach if the bootstrap estimation is accurate. Thirdly, our process for handling the non-positive definiteness and choosing a critical value may benefit from projection to a particular form of a covariance matrix that would lead to a more direct critical value choice. We could use theory on a specific type of covariance matrix, such as assuming an autoregressive covariance matrix. The addition of smoothing along the diagonal of the covariance matrix, while it improves the summary metrics, also introduces a bias and slight over-coverage to the right tail. We could also consider an off-diagonal correction after smoothing.

\vskip 1cm

\appendix

\section*{Appendix 1}
\subsection*{Proofs}

\begin{proof}[of Theorem~\ref{thm:covforU}]
For the fixed $t_1$ and $t_2$, the covariance of an order-$k$ complete U-statistics is
\begin{eqnarray}\label{eq:covofu}
  \text{cov}\lbrace U(t_1),\,U(t_2)\rbrace
  =&\binom{n}{k}^{-1}\binom{n}{k}^{-1}\sum_{S_i\subset\cD_n}\sum_{S_j\subset\cD_n}\text{cov}\Big\lbrace h(t_1;\,S_{i}),\,h(t_2;\,S_{j})\Big\rbrace \nonumber\\
  =&\binom{n}{k}^{-2}\sum_{d=1}^k\sum_{\mid S_{i}\cap S_{j}\mid =d}\text{cov}\Big\lbrace h(t_1;\,S_{i}),\,h(t_2;\,S_{j})\Big\rbrace.\nonumber
\end{eqnarray}
For a pairs of subsamples $S_i$ and $S_j$ with $d$ overlapping
samples, let
\[\xi_{d,\,k}^2(t_1,\,t_2)=\text{cov} \Big\lbrace h(t_1;\,S_i),\, h(t_2;\,S_j) \Big\rbrace.\]
Then we have
\begin{align*}
\xi_{d,\,k}^2(t_1,\,t_2)
=&\text{cov}\Big[ E \lbrace h(t_1;\,S) \mid  Z_{(1:d)} \rbrace,\,E \lbrace h(t_2;\,S) \mid  Z_{(1:d)} \rbrace \Big]\\
=&\text{cov}\Big\lbrace  h(t_1;\,S),\, h(t_2;\,S)\Big\rbrace - E\Big[ \text{cov} \lbrace h(t_1;\,S),\,h(t_2;\,S)  \mid  Z_{(1:d)} \rbrace\Big].
\end{align*}
Denote \[\tilde\xi_{d,\,k}^2(t_1,\,t_2)=E\Big[ \text{cov} \lbrace h(t_1;\,S),\,h(t_2;\,S)  \mid  Z_{(1:d)} \rbrace\Big],\]
then
\[\xi_{d,\,k}^2(t_1,\,t_2)=\text{cov}\Big\lbrace  h(t_1;\,S),\, h(t_2;\,S)\Big\rbrace-\tilde\xi_{d,\,k}^2(t_1,\,t_2),\]
hence
\begin{align*}
\text{cov}\lbrace U(t_1),\,U(t_2)\rbrace
=&\binom{n}{k}^{-2}\sum_{d=1}^k\sum_{\mid S_{i}\cap S_{j}\mid =d}\xi_{d,\,k}^2(t_1,\,t_2)\\
=&\binom{n}{k}^{-2}\sum_{d=1}^k\binom{n}{k}\binom{k}{d}\binom{n-k}{k-d}\xi_{d,\,k}^2(t_1,\,t_2)\\
=&\binom{n}{k}^{-1}\sum_{d=1}^k\binom{k}{d}\binom{n-k}{k-d}\left[\text{cov}\big\lbrace  h(t_1;\,S),\, h(t_2;\,S)\big\rbrace - \tilde\xi_{d,\,k}^2(t_1,\,t_2)\right]\\
=&\sum_{d=1}^k\gamma_{d,k,n}\left[\text{cov}\big\lbrace  h(t_1;\,S),\, h(t_2;\,S)\big\rbrace - \tilde\xi_{d,\,k}^2(t_1,\,t_2)\right]\\
=&(1-\gamma_{0,k,n})\,\text{cov}\big\lbrace  h(t_1;\,S),\, h(t_2;\,S)\big\rbrace -\\
&\left\lbrace\sum_{d=0}^k\gamma_{d,k,n}\,\tilde\xi_{d,\,k}^2(t_1,\,t_2)-\gamma_{0,k,n}\,\tilde\xi_{0,\,k}^2(t_1,\,t_2)\right\rbrace\\
=&\text{cov}\big\lbrace  h(t_1;\,S),\, h(t_2;\,S)\big\rbrace - \sum_{d=0}^k\gamma_{d,k,n}\,\tilde\xi_{d,\,k}^2(t_1,\,t_2)\\
=&C^{(h)}(t_1,t_2)-C^{(S)}(t_1,\,t_2).
\end{align*}

To estimate $C^{(h)}(t_1,\,t_2)$, we use the fact that
\bal
C^{(h)}(t_1,\,t_2)=&\text{cov}\Big\lbrace h(t_1;\,S),\,h(t_2;\,S)\Big\rbrace\nonumber\\
=&E\left[\frac{1}{2}\lbrace h(t_1;\,S_i)-h(t_1;\,S_j)\rbrace\lbrace h(t_2;\,S_i)-h(t_2;\,S_j)\rbrace\right]\nonumber
\eal
where the pair of $S_i$ and $S_j$ does not contain any overlapping samples,
hence $C^{(h)}(t_1,t_2)$ can also be estimated as
\bal
\hat{C}^{(h)}(t_1,\,t_2)=\frac{1}{2N_0}\sum_{S_i\cap S_j=\emptyset}
\Big\lbrace h(t_1;\,S_i)-h(t_1;\,S_j)\Big\rbrace\Big\lbrace h(t_2;\,S_i)-h(t_2;\,S_j)\Big\rbrace.\nonumber
\eal

To estimate $C^{(S)}(t_1,\,t_2)$, we need to deal with $\text{cov} \lbrace h(t_1;\,S),\,h(t_2;\,S)  \mid  Z_{(1:d)} \rbrace$ first. For a pair of subsamples $S_i$ and $S_j$ such that $S_i\cap S_j=Z_{(1:d)}$,
\[\text{cov} \big\lbrace h(t_1;\,S),\,h(t_2;\,S) \mid Z_{(1:d)}\big\rbrace
=E\left[\frac{1}{2}\lbrace h(t_1;\,S_i)-h(t_1;\,S_j)\rbrace\lbrace h(t_2;\,S_i)-h(t_2;\,S_j)\rbrace\right],\]
hence $\tilde{\xi}_{d,\,k}^2(t_1,\,t_2)$ can be estimated by all such pairs of subsamples,
\[\hat{\tilde{\xi}}_{d,\,k}^2(t_1,\,t_2)=\frac{1}{N_d}\sum_{\mid S_i\cap S_j\mid =d}\frac{1}{2}\lbrace h(t_1;\,S_i)-h(t_1;\,S_j)\rbrace\lbrace h(t_2;\,S_i)-h(t_2;\,S_j)\rbrace,\]
where
\[N_d=\binom{n}{k}\binom{k}{d} \binom{n-k}{k-d}\]
is the number of the pairs with $d$ overlapping samples. Then
\begin{align*}
\hat{C}^{(S)}(t_1,\,t_2)=&\sum_{d=0}^k\gamma_{d,k,n}\,\tilde\xi_{d,\,k}^2(t_1,\,t_2)\\
=&\sum_{d = 0}^k \gamma_{d,k,n} \left[\frac{1}{N_d}\sum_{\mid S_i\cap S_j\mid =d}\frac{1}{2}\,\lbrace h(t_1;\,S_i)-h(t_1;\,S_j)\rbrace\lbrace h(t_2;\,S_i)-h(t_2;\,S_j)\rbrace\right]\\
=&\sum_{d = 0}^k \binom{n}{k}^{-2}\sum_{\mid S_i\cap S_j\mid =d}\frac{1}{2}\,\lbrace h(t_1;\,S_i)-h(t_1;\,S_j)\rbrace\lbrace h(t_2;\,S_i)-h(t_2;\,S_j)\rbrace\\
=&\underbrace{\binom{n}{k}^{-2}}_{\text{number of pairs}}  \underbrace{\sum_{d = 0}^k \sum_{\mid S_i\cap S_j\mid =d}}_{\text{all pairs of $k$ samples}}\frac{1}{2}\,\lbrace h(t_1;\,S_i)-h(t_1;\,S_j)\rbrace\lbrace h(t_2;\,S_i)-\\
&h(t_2;\,S_j)\rbrace\\
=&\binom{n}{k}^{-2} \sum_{S_i\neq S_j}\frac{1}{2}\,\lbrace h(t_1;\,S_i)-h(t_1;\,S_j)\rbrace\lbrace h(t_2;\,S_i)-h(t_2;\,S_j)\rbrace.
\end{align*}
This term can be estimated  by using all of the $k$ data subsample $S_i$,
\begin{align*}
\hat{C}^{(S)}(t_1,\,t_2)=\frac{1}{N}\sum_{S_i\subset\cD_n}\lbrace h(t_1;\,S_i)-U(t_1)\rbrace\lbrace h(t_2;\,S_i)-U(t_2)\rbrace.
\end{align*}
\end{proof}

\begin{proof}[of Theorem~\ref{thm:biasUB}]

For given $t_1$ and $t_2$, the difference between $\text{cov}\lbrace U_{B}(t_1),\,U_{B}(t_2)\rbrace$ and its complete counterpart $\text{cov}\lbrace U(t_1),\,U(t_2)\rbrace$ is
\begin{align}
\label{eq:evar_ub}
E\Big[\text{cov}\lbrace U_{B}(t_1),\,U_B(t_2)\mid \cD_n\rbrace\Big]
=& E\Big[E\lbrace U_{B}(t_1)U_{B}(t_2)\mid \cD_n\rbrace-E\lbrace U_{B}(t_1)\mid \cD_n\rbrace E\lbrace U_{B}(t_2)\mid \cD_n\rbrace\Big]\nonumber\\
=& E\Big[E\lbrace U_{B}(t_1)U_{B}(t_2)\mid \cD_n\rbrace-U(t_1)U(t_2)\Big].
\end{align}

The first item in the above equation is
\begin{align*}
E\lbrace U_{B}(t_1)&U_{B}(t_2)\mid \cD_n\rbrace\\
=&E\left(\frac{1}{4B^2}\left[\sum_{b=1}^B\lbrace h(t_1;\,S_{b1})+h(t_1;\,S_{b2})\rbrace\right]\left[\sum_{b=1}^B\lbrace h(t_2;\,S_{b1})+h(t_2;\,S_{b2})\rbrace\right]\mid \cD_n\right)\\
=&\frac{1}{4B^2}\sum_{b=1}^BE\left[\left\lbrace h(t_1;\,S_{b1})+h(t_1;\,S_{b2})\right\rbrace\left\lbrace h(t_2;\,S_{b1})+h(t_2;\,S_{b2})\right\rbrace\mid \cD_n\right] \\
& + \frac{2}{4B^2}\sum_{b1<b2}E\left[\left\lbrace h(t_1;\,S_{b1,1})+h(t_1;\,S_{b1,2})\right\rbrace\left\lbrace h(t_2;\,S_{b2,1})+h(t_2;\,S_{b2,2})\right\rbrace\mid \cD_n\right]\\
&\doteq\text{Part I}+\text{Part II}.
\end{align*}
We have
\begin{align*}
\text{Part I} =& \frac{1}{4B^2}\sum_{b=1}^BE\Big\lbrace h(t_1;\,S_{b1})h(t_2;\,S_{b1}) + h(t_1;\,S_{b2})h(t_2;\,S_{b2})\\
& \qquad\qquad\qquad + h(t_1;\,S_{b1}) h(t_2;\,S_{b2}) + h(t_1;\,S_{b2}) h(t_2;\,S_{b1})\mid  \cD_n\Big\rbrace\\
=& \frac{2}{4B^2}\sum_{b=1}^BE\left\lbrace h(t_1;\,S_{b1})h(t_2;\,S_{b1}) \mid  \cD_n\right\rbrace + \frac{2}{4B^2}\sum_{b=1}^BE\left\lbrace h(t_1;\,S_{b1}) h(t_2;\,S_{b2}) \mid  \cD_n\right\rbrace\\
=& \frac{1}{2\,B\,N}\sum_{S_i\subset\cD_n} h(t_1;\,S_{i})h(t_2;\,S_{i}) + \frac{1}{2\,B\,N\,\binom{n-k}{k}}\sum_{S_i\cap S_j =\emptyset}h(t_1;\,S_{i})\,h(t_2;\,S_{j}).
\end{align*}
Since
\begin{align*}
 & \sum_{S_i\cap S_j =\emptyset}\left\lbrace h(t_1;\,S_{i})+h(t_1;\,S_{j})\right\rbrace \left\lbrace h(t_2;\,S_{i})+h(t_2;\,S_{j})\right\rbrace\\
 =& \binom{n-k}{k}\sum_{S_i\subset\cD_n} 2\,h(t_1;\,S_{i})h(t_2;\,S_{i})
 + 2\sum_{S_i\cap S_j =\emptyset}h(t_1;\,S_{i})h(t_2;\,S_{j}),
\end{align*}
we get
\begin{align*}
\text{Part I}
 =&  \frac{1}{2BN}\sum_{S_i\subset\cD_n} h(t_1;\,S_{i})\,h(t_2;\,S_{i}) -\frac{1}{2BN}\sum_{S_i\subset\cD_n}h(t_1;\,S_{i})h(t_2;\,S_{j})\\
 & \quad + \frac{1}{4BN\binom{n-k}{k}}\sum_{S_i\cap S_j =\emptyset}\left\lbrace h(t_1;\,S_{i})-h(t_1;\,S_{j})\right\rbrace \left\lbrace h(t_2;\,S_{i})-h(t_2;\,S_{j})\right\rbrace\\
=&  \frac{1}{4BN\binom{n-k}{k}}\sum_{S_i\cap S_j =\emptyset}\left\lbrace h(t_1;\,S_{i}) + h(t_1;\,S_{j})\right\rbrace \left\lbrace h(t_2;\,S_{i}) + h(t_2;\,S_{j})\right\rbrace .
\end{align*}

The second term of (\ref{eq:evar_ub}) is
\begin{align*}
\text{Part II}
=&\frac{1}{B^2}\sum_{1\leq b2<b2\leq B}E\left[ h(t_1;\,S_{b1,1})\left\lbrace h(t_2;\,S_{b2,1})+h(t_2;\,S_{b2,2})\right\rbrace\mid  \cD_n\right]\\
=&\frac{B-1}{2B}\,E\left[ h(t_1;\,S_{11})\left\lbrace h(t_2;\,S_{21})+h(t_2;\,S_{22})\right\rbrace\mid  \cD_n\right]\\
=&\frac{B-1}{2B}\frac{1}{N^2\binom{n-k}{k}}\sum_{\genfrac{}{}{0pt}{}{i,\,j,\,j'}{S_j\cap S_{j'}=\emptyset}}h(t_1;\,S_{i})\left\lbrace h(t_2;\,S_{j})+h(t_2;\,S_{j'})\right\rbrace\\
=&\frac{B-1}{2B}\frac{1}{N^2\binom{n-k}{k}} \lbrace\sum_{i,\,j}\sum_{j':\,S_j\cap S_{j'}=\emptyset}\,h(t_1;\,S_{i})h(t_2;\,S_{j})\\
&+ \sum_{i,\,j'}\sum_{j:\,S_j\cap S_{j'}=\emptyset}\,h(t_1;\,S_{i})h(t_2;\,S_{j'})\rbrace\\
=&\frac{B-1}{2B}\frac{1}{N^2} \left\lbrace\sum_{i,\,j}\,h(t_1;\,S_{i})h(t_2;\,S_{j}) + \sum_{i,\,j'}\,h(t_1;\,S_{i})h(t_2;\,S_{j'})\right\rbrace\\
=&\left(1-\frac{1}{B}\right)\left\lbrace\frac{1}{N}\sum_{S_i\subset\cD_n} h(t_1;\,S_{i})\right\rbrace\left\lbrace\frac{1}{N}\sum_{S_i\subset\cD_n} h(t_2;\,S_{i})\right\rbrace
\end{align*}

Hence
\begin{align}\label{eq:EUBsquare}
\text{cov}\lbrace U_{B}(t_1),&\,U_B(t_2)\mid \cD_n\rbrace=\text{Part I} + \text{Part II} - U(t_1)\,U(t_2)\\
=& \frac{1}{4BN\binom{n-k}{k}}\sum_{S_i\cap S_j =\emptyset}\left\lbrace h(t_1;\,S_{i})+h(t_1;\,S_{j})\right\rbrace \left\lbrace h(t_2;\,S_{i})+h(t_2;\,S_{j})\right\rbrace\nonumber\\
&-\frac{1}{B}\,U(t_1)\,U(t_2)\nonumber
\end{align}
Therefore the difference between $\text{cov}\lbrace U_{B}(t_1),\,U_{B}(t_2)\rbrace$ and $\text{cov}\lbrace U(t_1),\,U(t_2)\rbrace$ is
\begin{align*}
E\Big[\text{cov}\lbrace U_{B}(t_1),\,U_B(t_2)\mid \cD_n\rbrace\Big]
=& \frac{1}{4B}\, E [\lbrace h(t_1;\,S_{i})+h(t_1;\rbrace \lbrace h(t_2;\,S_{i})\\
&+h(t_2;\,S_{j})\rbrace]-\frac{1}{B}\,E\lbrace\,U(t_1)U(t_2)\rbrace \\
=& \frac{1}{2B}\Big[ E\lbrace\,h(t_1;\,S)h(t_2;\,S)\rbrace + E\lbrace\,h(t_1;\,S)\,E\,h(t_2;\,S)\rbrace\\
&- 2\,E\lbrace\,U(t_1)U(t_2)\rbrace\Big]\\
=& \frac{1}{2B}\Big[ \text{cov}\lbrace h(t_1;\,S),\,h(t_2;\,S)\rbrace - 2\,\text{cov}\lbrace U(t_1),\,U(t_2)\rbrace\Big]\\
=&\frac{1}{2B}C^{(h)}(t_1,\,t_2)-\frac{1}{B}\text{cov}\lbrace U(t_1),\,U(t_2)\rbrace.
\end{align*}
Hence we get the result.
\end{proof}

\begin{proof}[of Theorem~\ref{thm:biasVS}]
Since
\begin{align*}
\hat{C}^{(S)}_{B}(t_1,\,t_2)
=& \frac{1}{2B}\sum_{b=1}^B\Big[ h(t_1;\,S_{b1})h(t_2;\,S_{b1})+h(t_1;\,S_{b2})h(t_2;\,S_{b2}) + 2U_{B}(t_1)U_{B}(t_2)\\
& \qquad\qquad - 2\lbrace h(t_2;\,S_{b1})+h(t_2;\,S_{b2})\rbrace U_B(t_1) - 2\lbrace h(t_1;\,S_{b1})\\
& \qquad\qquad +h(t_1;\,S_{b2})\rbrace U_B(t_2) \Big]\\
=& \frac{1}{2B}\sum_{b=1}^B\Big\lbrace h(t_1;\,S_{b1})h(t_2;\,S_{b1})+h(t_1;\,S_{b2})h(t_2;\,S_{b2})\Big\rbrace - U_{B}(t_1)U_B(t_2)
\end{align*}
together with \eqref{eq:EUBsquare}, we have
\begin{align*}
E\left\lbrace\hat{C}^{(S)}_{B}(t_1,\,t_2)\right\rbrace
=& E \left[ \frac{1}{B}\sum_{b=1}^B E \left\lbrace  h(t_1;\,S_{b1})h(t_2;\,S_{b1}) \mid  \cD_n\right\rbrace\right]\\
&- E \Big[ \text{cov}\lbrace U_{B}(t_1),\,U_B(t_2)\mid \cD_n\rbrace + U(t_1)U(t_2)\Big]\\
=&E \left\lbrace\frac{1}{N}\sum_{S_i\subset\cD_n} h(t_1;\,S_{i}) h(t_2;\,S_{i})- U(t_1)U(t_2) \right\rbrace\\
&- \frac{1}{B}\left[ \frac{1}{2}\,C^{(h)}(t_1,\,t_2)-\text{cov}\lbrace U(t_1),\,U(t_2)\rbrace\right]
 \\
=& \left[C^{(h)}(t_1,\,t_2)- \text{cov}\lbrace U(t_1),\,U(t_2)\rbrace \right] \\
&- \frac{1}{B}\left[ \frac{1}{2}\,C^{(h)}(t_1,\,t_2)-\text{cov}\lbrace U(t_1),\,U(t_2)\rbrace\right]\\
=& \left( 1-\frac{1}{B}\right)C^{(S)}(t_1,\,t_2) + \frac{1}{2B}\,C^{(h)}(t_1,\,t_2).
\end{align*}

\end{proof}

\section*{Appendix 2}
\subsection*{Additional Figures}

\begin{figure}[ht]
    \centering
    \includegraphics{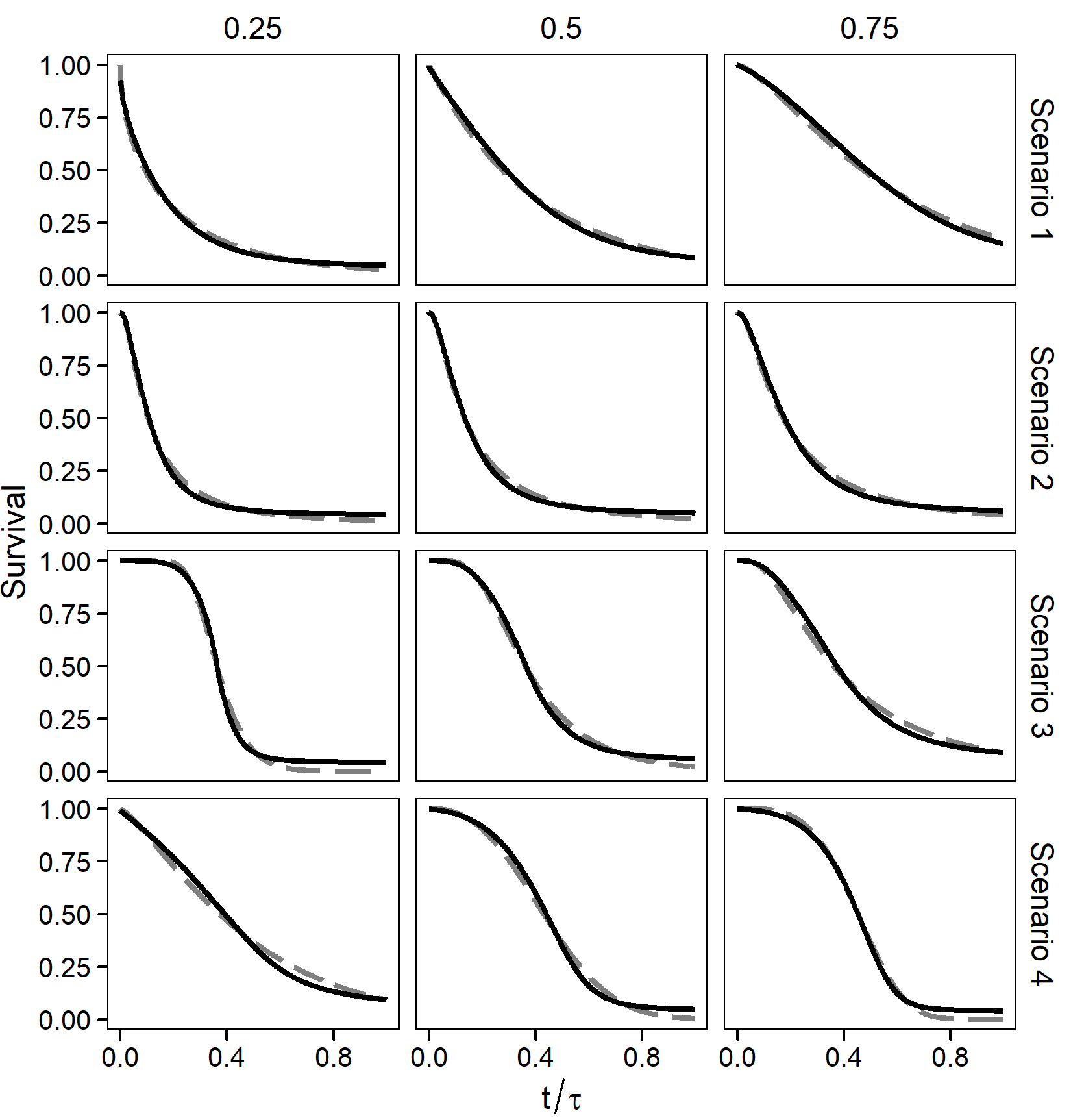}
    \caption{Averaged survival prediction (grey dashes) versus the true lifetime distributions (black solid).}
    \label{fig:avg_true}
\end{figure}

\begin{figure}[ht]
    \centering
    \includegraphics[width=0.95\textwidth]{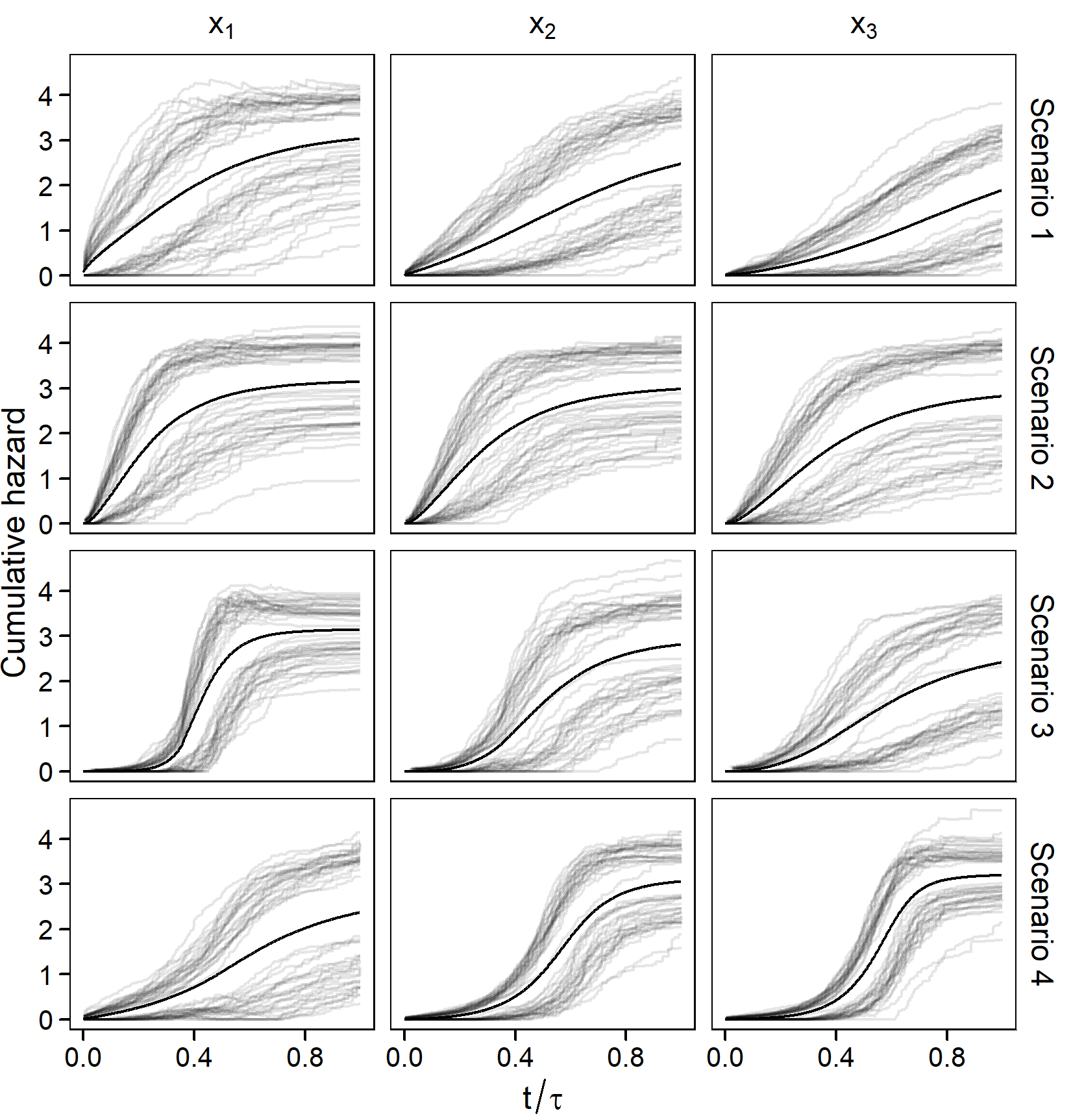}
    \caption{ For each scenario, 25 confidence bands constructed by smoothed projection method (grey) and the true cumulative hazard function (black).}
    \label{fig:manyCB}
\end{figure}

\begin{figure}[ht]
    \centering
    \includegraphics{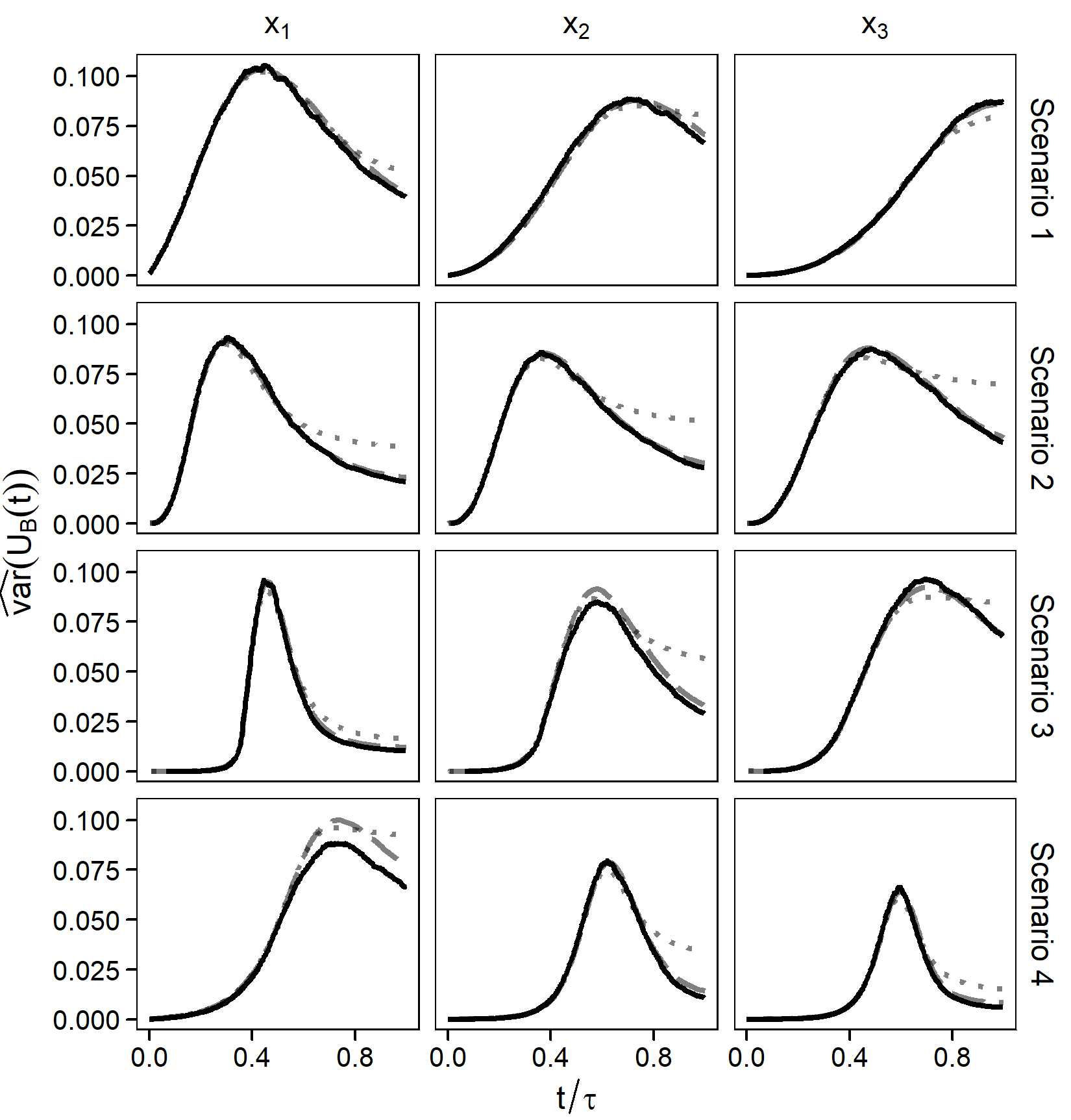}
    \caption{A comparison of the average estimated variances calculated by the original projection method (dashed grey) and the smoothed projection method (dotted grey) with the true variance (solid black).}
    \label{fig:var_avg}
\end{figure}

\begin{table}
\caption{D-optimality values for design $X$ under five different scenarios.  \label{tab:tabone}}
\begin{center}
\begin{tabular}{rrrrr}
one & two & three & four & five\\\hline
1.23 & 3.45 & 5.00 & 1.21 & 3.41 \\
1.23 & 3.45 & 5.00 & 1.21 & 3.42 \\
1.23 & 3.45 & 5.00 & 1.21 & 3.43 \\
\end{tabular}
\end{center}
\end{table}

\bibliographystyle{Chicago}
\bibliography{ref}
\end{document}